\documentclass[twocolumn]{aastex631}

\newcommand{\ang}{$\rm\AA$}

\newcommand{\lha}{L$_{H\alpha}$}

\newcommand{\hb}{H$\beta$}

\newcommand{\hii}{H\,{\sc II} }
\newcommand{\hi}{H\,{\sc I} }
\newcommand{\oii}{[O{\sc II}]}
\newcommand{\oiii}{[O{\sc III}]}
\newcommand{\nii}{[N{\sc II}]}
\newcommand{\sii}{[S{\sc II}]}
\newcommand{\siii}{[S{\sc III}]}

\defcitealias{Cosens2022}{Paper 1}
\defcitealias{KK04}{KK04}
\defcitealias{PT05}{PT05}

\usepackage{amsmath}

\begin{document}

\title{Oxygen Abundance Throughout the Dwarf Starburst IC\,10}

\correspondingauthor{Maren Cosens}
\email{mcosens@carnegiescience.edu}

\author[0000-0002-2248-6107]{Maren Cosens}
\affiliation{The Observatories, Carnegie Science, 813 Santa Barbara Street, Pasadena, CA 91101} 

\author[0000-0003-1034-8054]{Shelley A. Wright}
\affiliation{Physics Department, University of California, San Diego, 9500 Gilman Drive, La Jolla, CA 92093 USA} 
\affiliation{Department of Astronomy, University of California, San Diego, 9500 Gilman Drive, La Jolla, CA 92093 USA}

\author[0000-0002-4378-8534]{Karin Sandstrom}
\affiliation{Department of Astronomy, University of California, San Diego, 9500 Gilman Drive, La Jolla, CA 92093 USA}

\author[0000-0003-3498-2973]{Lee Armus}
\affiliation{Spitzer Science Center, California Institute of Technology, 1200 E. California Blvd., Pasadena, CA 91125 USA}

\author[0000-0002-8659-3729]{Norman Murray}
\affiliation{Canadian Institute for Theoretical Astrophysics, University of Toronto, 60 St. George Street, Toronto, ON M5S 3H8, Canada} 

\author[0000-0003-4852-8958]{Jordan N. Runco}
\affiliation{Department of Physics \& Astronomy, University of California, Los Angeles, Los Angeles, CA 90095, USA}

\author[0000-0002-8780-8226]{Sanchit Sabhlok}
\affiliation{Physics Department, University of California, San Diego, 9500 Gilman Drive, La Jolla, CA 92093 USA} 

\author[0000-0003-2687-9618]{James Wiley}
\affiliation{Physics Department, University of California, San Diego, 9500 Gilman Drive, La Jolla, CA 92093 USA} 

\begin{abstract}
Measurements of oxygen abundance throughout galaxies provide insight to the formation histories and ongoing processes. Here we present a study of the gas phase oxygen abundance in the \hii regions and diffuse gas of the nearby starburst dwarf galaxy, IC\,10. Using the Keck Cosmic Web Imager (KCWI) at W.~M.~Keck Observatory, we map the central region of IC\,10 from 3500-5500\ang. The auroral \oiii4363\ang \, line is detected with high signal-to-noise in 12 of 46 \hii regions observed, allowing for direct measurement of the oxygen abundance, yielding a median and standard deviation of $\rm12+log(O/H)=8.37\pm0.25$. We investigate trends between these directly measured oxygen abundances and other \hii region properties, finding weak negative correlations with the radius, velocity dispersion, and luminosity. We also find weak negative correlations between oxygen abundance and the derived quantities of turbulent pressure and ionized gas mass, and a moderate correlation with the derived dynamical mass. Strong line, $\rm R_{23}$ abundance estimates are used in the remainder of the \hii regions and on a resolved spaxel-by-spaxel basis. There is a large offset between the abundances measured with $\rm R_{23}$ and the auroral line method. We find that the $\rm R_{23}$ method is unable to capture the large range of abundances observed via the auroral line measurements. The extent of this variation in measured abundances further indicates a poorly mixed interstellar medium (ISM) in IC\,10, which is not typical of dwarf galaxies and may be partly due to the ongoing starburst, accretion of pristine gas, or a late stage merger.
\end{abstract}

\section{Introduction} \label{sec:intro}
Processes within galaxies such as their star formation history, inflows and outflows, turbulent mixing, and merger history all leave an imprint on the quantity and distribution of metals in the interstellar medium (ISM) \citep{DiMatteo2009, Dave2011, Dayal2013, Petit2015, KrumholzTing2018}. This makes measuring the oxygen abundance throughout a galaxy an important tool in understanding how these processes play out in different environments and influence the evolution of the galaxy.

Spiral and elliptical galaxies are consistently found to have metallicity gradients, with decreasing metallicity as a function of increasing radius \citep[e.g.,][]{Smith1975, Zaritsky1994, Werk2011, Kreckel2019} typically attributed to inside-out galaxy growth \citep[e.g.,][]{Boissier1999}. Galaxies in the early universe, on the other hand, show significantly more variation than local spirals and ellipticals, but for the most part exhibit very shallow radial metallicity gradients if one is found at all \citep[e.g.,][and references therein]{Wuyts2016, Wang2017, ForsterSchreiber2018, Maiolino2019}. This is often attributed to these systems being out of equilibrium and more sensitive to recent gas accretion or bursts of star formation, with simulations showing significant fluctuation of abundance gradients with time \citep[e.g.,][]{Ma2017}. 

Since we cannot observe these high-z systems with the same tracers and in the same amount of detail as local systems, one approach is to find local ``analogues" that \textit{can} be studied in detail. Dwarf galaxies are typically used for this purpose as their ``low" metallicities ($\rm \lesssim0.5Z_{\sun}$) are more similar to early universe environments. Unlike spiral galaxies, dwarf galaxies generally are not observed to have metallicity gradients and are instead considered to be well mixed \citep{Berg2012}. This is generally attributed to efficient mixing with recent simulations highlighting the larger impact of feedback on low mass galaxies \citep[e.g.,][]{Ma2017, Mercado2020}. There are also a handful of dwarf galaxies ($\rm \lesssim10$) observed to be inhomogenous with explanations invoking a temporary disturbance such as merger activity or bursty star formation episodes \citep[see discussion in][]{James2020, Fernandez-Arenas2023, Gao2023}.

Integral field spectrographs (IFS) have been a useful tool for studying the abundance variations on increasingly resolved scales throughout all of these types of galaxies. With the ability to generate maps of emission lines and derived properties, the variation in abundance can be investigated in greater detail than with slit spectroscopy. 

IC\,10 is an ideal target for such a resolved study of oxygen abundance variations in a nearby dwarf galaxy. It is the nearest starburst to us \citep[$\rm715\,kpc$;][]{Kim2009} and is well resolved with over 140 \hii regions throughout the galaxy \citep{HL1990}. IC\,10 has a low global metallicity at only $\rm12+log(O/H)\sim8.2$ \citep[e.g.,][]{Skillman1989, Lebouteiller2012}, making it more analogous to galaxies in the early universe that evolve to present day spirals and ellipticals. The current starburst makes it particularly interesting to consider as an analogue to galaxies in the early universe in which stars were being formed much more rapidly than today. There is also a large \hi reservoir around IC\,10 with distinct kinematic features indicative of either ongoing accretion \citep[e.g.,][]{WilcotsMiller1998} or interaction with a possible companion galaxy \citep[e.g.,][]{Nidever2013, Ashley2014}. Both of these scenarios could result in variation of the chemical conditions throughout the ISM.

We previously studied the \hii regions in the central region of IC\,10 with the high resolution mode of the Keck Cosmic Web Imager \citep[KCWI; ][]{Morrissey2018} IFS in order to study their morphology, kinematics, and feedback processes \citep[][hereafter Paper 1]{Cosens2022}. We found that these \hii regions were offset from typical scaling relationships and the gas kinematics indicate that the regions are generally young and show evidence of ongoing, feedback-driven expansion.

We present here a follow-up study of the same \hii regions and diffuse gas in IC\,10 with a lower spatial and spectral resolution mode of KCWI in order to extend the wavelength coverage and improve detection of the faint emission lines necessary to study the gas phase oxygen abundance throughout this galaxy. The observations and data reduction are described in Section \ref{sec:obs}. In Section \ref{sec:metallicity} we determine the gas phase oxygen abundance from the integrated and spatially resolved spectra via the direct and strong line methods. We discuss these results in Section \ref{sec:discussion} with the level of ISM mixing in IC\,10 (Section \ref{sec:mixing}), trends between the oxygen abundance and other \hii region properties (Section \ref{sec:prop_trends}), Wolf-Rayet (WR) star locations (Section \ref{sec:WRstars}), and finally, potential sources of systematic uncertainty in this analysis (Section \ref{sec:unc}). We summarize the results in Section \ref{sec:conclusion}.

\section{Observations \& Data Reduction} \label{sec:obs}
\begin{figure}
    \centering
    \includegraphics[width=0.5\textwidth]{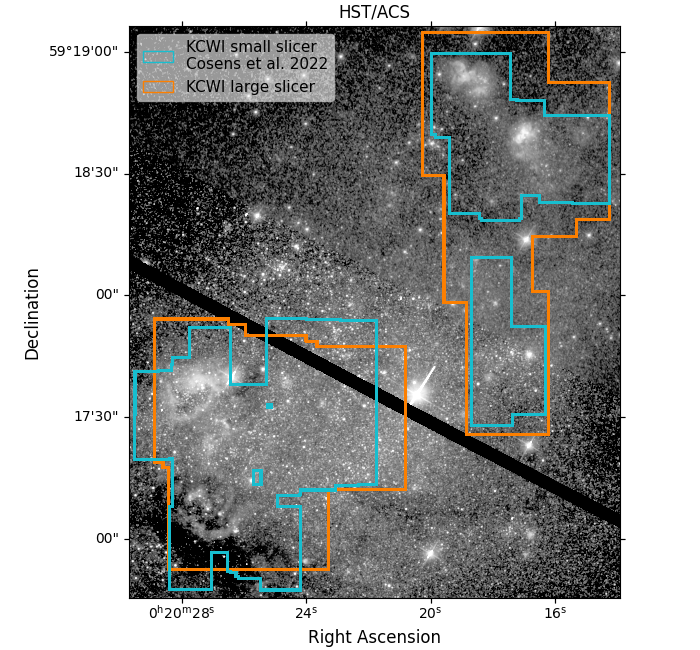}
    \caption{KCWI footprint overlayed on HST/ACS image of the central region of IC\,10 taken with the F606W filter. The footprint of the previous high resolution KCWI observations discussed in \citetalias{Cosens2022} is shown in blue, while the new observations are outlined in orange.}
    \label{fig:obs_footprint}
\end{figure}

We observed IC\,10 with the KCWI IFS at the W.M. Keck Observatory on the nights of August 13, 2021 and November 25, 2021. These observations made use of the large slicer and low resolution BL grating available on KCWI. This configuration resulted in a spatial sampling of 1.35$\arcsec$/pixel and spectral resolving power R$\sim$900 over a wavelength range of 3500 - 5500\ang. The large slicer provides a 33$\arcsec\times$20.4$\arcsec$ field of view (FoV). Our observations covered 9 unique pointings of IC\,10 with 4$\times$300s exposures at each one with a dither pattern of 0, -0.5, -1, +2 slices\footnote{(-): left; (+): right}. Sky frames were taken approximately every 45 minutes by pointing to a clear patch of sky offset from the extended diffuse emission of IC\,10. At each new exposure the associated guider image was saved to be used for WCS verification and possible correction. The footprint of these KCWI observations is shown in Figure \ref{fig:obs_footprint}; these fields were chosen to encompass many of the bright \hii regions in the central portion of IC\,10, as well as to match our previous observations with the higher resolution mode of KCWI in \citetalias{Cosens2022}. These complimentary observations made use of the small slicer (0.35$\arcsec$/pixel) and BH3 grating (R$\sim$18,000; 4700-5200\ang) to study the detailed structure and kinematics of IC\,10's \hii regions, but did not cover the bluer wavelengths necessary to obtain abundance estimates throughout the ionized gas.

Observations were reduced using the publicly available KCWI Data Reduction Pipeline (DRP) v1.1.0 written in IDL \citep{KCWIDRP} with some modifications and additions for this data set. These changes are described in more detail in \citetalias{Cosens2022} and summarized here. The bias subtraction performed as part of DRP stage 1 is modified to be a scaled bias subtraction, using the overscan regions of the master bias and science frame to determine the scaling factor for each row of the detector. We skip the default sky subtraction in stage 5 of the DRP in favor of our own scaled sky subtraction using the final data cubes in order to better account for differences in the sky background levels over time. Finally we correct for WCS errors between frames and observing nights by aligning each frame to the coordinates of the HST/ACS imaging of IC\,10 as a baseline reference. These WCS offsets are corrected by measuring the positions of bright stars in the guider frames associated with each pointing and determining the average offset from the HST imaging. This offset was on average $\rm <1$\,pixel in declination and $\sim \rm 2$\,pixels in right ascension. All science frames were then mosaicked using the Python package \texttt{reproject} \citep{reproject}. Pixels were also re-binned from rectangular to square during this step.

\subsection{Spectral Extraction} \label{sec:spec_extraction}
The 46 \hii regions in the field were identified in \citetalias{Cosens2022} using the high resolution mode KCWI observations and the python package \texttt{astrodendro}. The identification procedure is summarized briefly here. First, preliminary flux maps were constructed by integrating over the location of the \oiii5007\ang \, line and performing a conservative subtraction of the Diffuse Ionized Gas (DIG). This DIG subtraction simply used the flux measured in an area of the field with no known or apparent \hii regions in order to reduce confusion with filamentary and diffuse structure in the region identification procedure. \texttt{astrodendro} was then run on these DIG corrected, 2D flux maps in order to identify the \hii regions with both built-in and custom constraints applied on what should be considered an independent region. This included the minimum peak value for a structure, the minimum flux required to include a pixel in a structure, the mimimun step size between independent regions, and custom constraints on the minimum radius and minor axis length of a structure. These values were determined from the variance cube produced by the KCWI DRP as well as measurements of the point spread function (PSF) in observed standard stars. A small number of narrow filaments are still identified with this procedure and are manually removed as well as any regions which are truncated by the edge of the FoV.

Three of the \hii regions identified in \citetalias{Cosens2022} were truncated in the coarse resolution observations and so are excluded here. For the rest, integrated \hii region spectra are extracted over a circular aperture with the center and radius determined in the \hii region identification routine. As discussed in \citetalias{Cosens2022}, there are multiple ways to define the characteristic radius of an \hii region and we use here a pseudo-halflight radius, $r_{1/2}^* = \sqrt{A_{\rm{ellipse}}/\pi}$, where $A_{\rm{ellipse}}$ is the area determined by \texttt{astrodendro} from the second moments of the region contours. The \texttt{aperture\_photometry} function in the \texttt{photutils} Python package is used with these centers and radii to extract integrated flux and error spectra for each region from the KCWI science and variance data cubes. In analysis which produces resolved maps, spectra are extracted over individual spaxels, with error spectra also being derived from the variance cube.

All spectra, resolved and integrated, are corrected for the contribution of underlying DIG before further analysis is performed. An average DIG spectrum is derived from the same region used for the preliminary subtraction of the \oiii5007\ang \, flux map before \hii region identification. Any wavelength shift between this DIG spectrum and the science spectra is corrected for by fitting the \oiii5007\ang \, lines with Gaussian profiles and calculating the difference in the line centers. The DIG spectrum is then subtracted from the science spectrum with a scaling factor for the number of spaxels in the integrated spectrum (1 for individual spaxel spectra; N pixels for integrated spectra) and the errors are propagated in the associated error spectrum. Finally, all spectra are then corrected for extinction based on the ratio of the H$\gamma$ and \hb \, emission lines. The reddening is calculated following \citet{Momcheva2013}'s Equation A10:
\begin{equation}
    \small
    \rm E(B-V) = \frac{-2.5}{\kappa(H\beta)-\kappa(H\gamma)}\times log_{10}\left(\frac{0.47}{\left(H\gamma / H\beta)\right )_{\rm{obs}}}\right)
\end{equation}
where $\kappa(H\beta)=4.6$ and $\kappa(H\gamma)=5.12$. The average reddening is $\rm E(B-V) = 0.85$ in the integrated \hii region spectra and $\rm E(B-V) = 0.81$ in the individual spaxel spectra. All spectra are corrected for reddening with the measured $\rm E(B-V)$ and a \citet{Cardelli1989} extinction law. The remainder of the analysis is performed on the DIG subtracted and extinction corrected spectra.

\begin{figure*}
    \centering
    \includegraphics[width=0.98\textwidth]{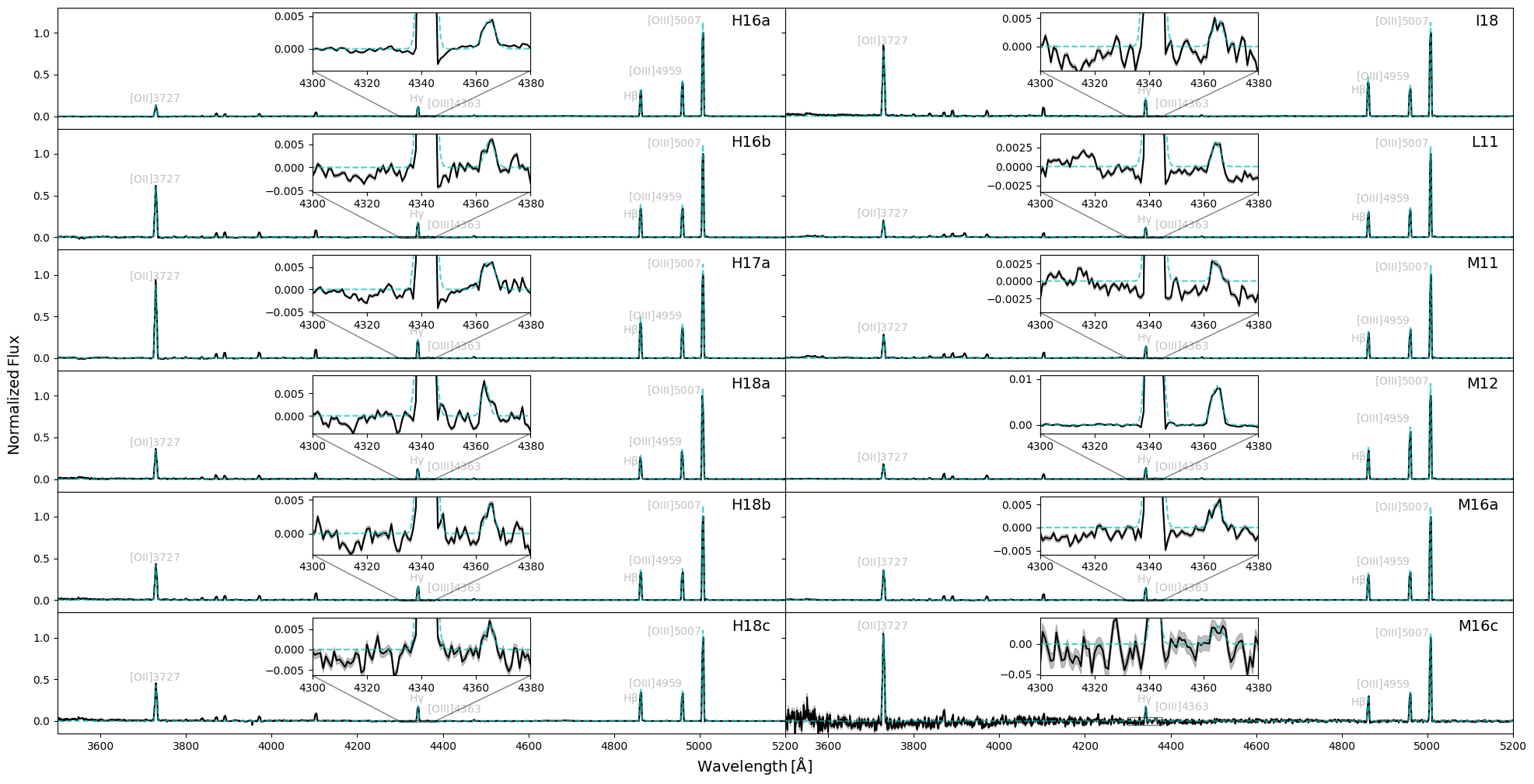}
    \caption{Integrated spectra of each \hii region with $\rm SNR_{\oiii4363}>3$, shown here after reddening correction, DIG subtraction, and continuum subtraction. The error spectra are shown by the gray shading while the emission line fits are shown with the dashed cyan line. All spectra are normalized to the \oiii5007\ang \, line for ease of comparison. The insets show a zoom-in of the fit to $\rm H\gamma$ and \oiii4363\ang\, to illustrate the clear detection of the faint auroral line. }
    \label{fig:spectra}
\end{figure*}

All emission line fits referred to in the following sections are single Gaussian fits performed after continuum subtraction. Each line is fit with independent parameters for the line center, width, and amplitude. Figure \ref{fig:spectra} shows the spectra and emission line fits for all \hii regions discussed in Section \ref{sec:direct}.

\section{Oxygen Abundance} \label{sec:metallicity}
The oxygen abundance within the \hii regions and surrounding gas is determined from both the direct and indirect methods. The direct method requires detection of a faint auroral line --- \oiii4363\ang \, in the wavelength range of KCWI --- in order to estimate the electron temperature of the gas. The indirect, or strong line, method on the other hand does not require detection of this faint emission line and can therefore be extended to fainter \hii regions and even resolved oxygen abundance estimates using the spectra of individual spaxels. However, these strong line methods rely on theoretical or empirical calibrations without direct information about gas densities and temperatures and therefore suffer from systematic uncertainties and variation between calibrations.

Throughout this section we will report only measurement uncertainties for individual values. Additional systematic uncertainties are discussed in Section \ref{sec:unc}

\subsection{Direct Metallicity} \label{sec:direct}
\begin{figure*}
    \gridline{\fig{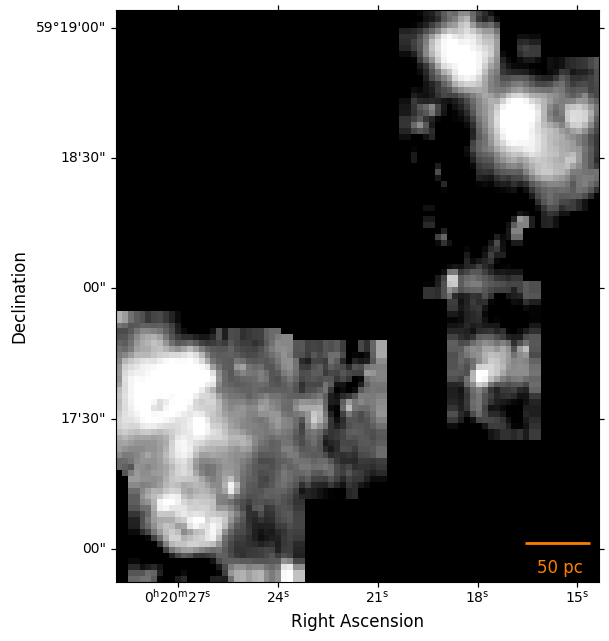}{0.37\textwidth}{(a)}}
    \gridline{\fig{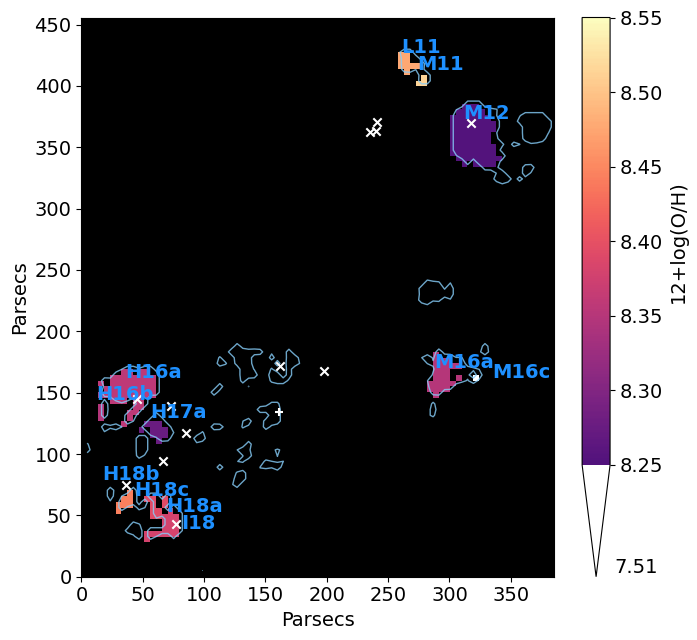}{0.43\textwidth}{(b)}
        \fig{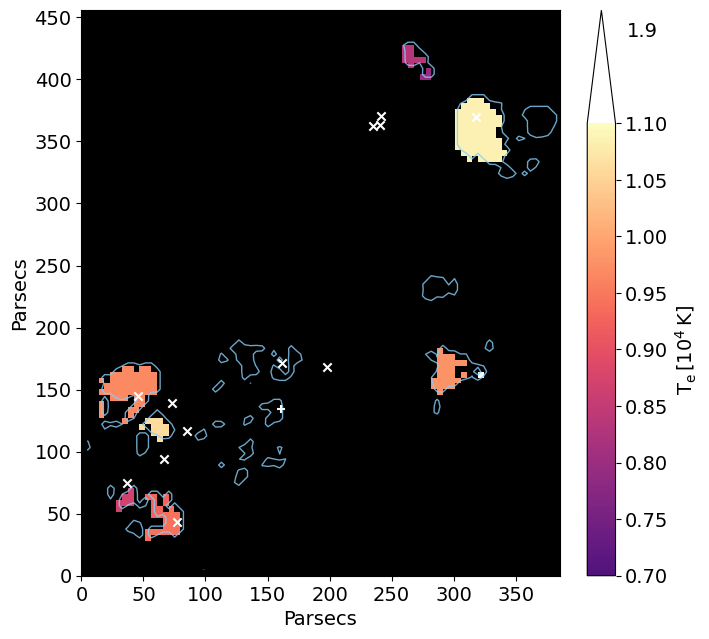}{0.43\textwidth}{(c)}}
    \caption{(a): Integrated \oiii5007\ang \, flux map utilizing the KCWI large slicer and BL grating with a 50\,pc scale bar at the bottom right. These observations cover the field shown by the orange contours in Figure \ref{fig:obs_footprint}. (b): Map of oxygen abundance in those 12 \hii regions with sufficient SNR to detect \oiii4363\ang \, required for the direct method ($\rm SNR_{\oiii4363}>3$). These detections are in the integrated region spectra so the color is shown filling the region contour. The remaining \hii regions are shown by the unfilled blue contours. The outlier region M16c -- discussed further in Section \ref{sec:discussion} -- is shown in white in order to increase the contrast of the variations seen between the other eleven regions which span a narrower range of oxygen abundance. These maps cover the exact same field as those in (a), but the axes are converted to pc to illustrate the spatial scale and separation of the \hii regions. (c): The electron temperature determined from the \oiii4364\ang \, line following Equation \ref{eqn:Te}. The locations of spectroscopically confirmed Wolf-Rayet (WR) stars are shown with white X's and unconfirmed with white $\rm +$'s. These maps illustrate the range of oxygen abundance accross \hii regions on even these small distances within the central region of IC\,10.}
    \label{fig:direct_metallicity}
\end{figure*}

The \oiii4363\ang \, auroral line is detected with SNR$>$3 in twelve of the integrated \hii region spectra, and therefore can be used to infer the gas phase oxygen abundance. This is calculated following the procedure laid out in \citet{Perez-Montero2017} in which the electron temperature, $\rm T_e$, is inferred from the ratio of the auroral and strong lines of the same species --- in this case \oiii5007\ang \, and \oiii4959\ang:

\begin{eqnarray}
    \rm T_e(10^4\,K) = 0.784 - 0.0001357\times R_{O3} +\frac{48.44}{R_{O3}} \label{eqn:Te}\\
    R_{O3} = \frac{F(\oiii5007\AA) + F(\oiii4959\AA)}{F(\oiii4363\AA)}
\end{eqnarray}
This equation is valid over the range $\rm 0.7 \leq T_e (10^4\,K) \leq 2.5$ with the measurements of all twelve \hii regions falling within this range. The abundance of the $\rm O^{2+}$ ion can be determined from this value of $\rm T_e$ as well as the flux of the \oiii4959,5007\ang \, and \hb \, lines:
\begin{multline}
    \rm 12+log(O^{2+}/H)=log \left( \frac{F(\oiii4959\AA)+F(\oiii5007\AA)}{F(H\beta)} \right) \\
    \rm +6.1868+\frac{1.2491}{T_e}-0.5816log(T_e)
\end{multline}
The median value and associated uncertainty in the \hii regions of our sample is $\rm 12+log(O^{2+}/H) = 8.05 \pm 0.18$. Unfortunately, the spectral resolution of the large slicer - BL mode of KCWI is not sufficient to deblend the \oii3727,3729\ang \, doublet, so we cannot directly probe the electron density to calculate the abundance of $\rm O^+$. In order to estimate the total oxygen abundance, therefore, we must use an empirical relationship between $\rm 12+log(O/H)$ and the \oiii \, electron temperature, $\rm T_e$ \citep{Amorin2015}:
\begin{equation}
    \rm 12+log(O/H)= (9.22\pm0.03) - (0.89\pm0.02)\times T_e
\end{equation}
The median total oxygen abundance determined in IC\,10's \hii regions with this method is $\rm12+log(O/H)=8.37\pm 0.18$, and is shown for each region in Figure \ref{fig:direct_metallicity}. We will use this total oxygen abundance in the remainder of the paper, making this a ``semi-direct" measurement.

This median value is consistent with previous studies of IC\,10 \hii regions and planetary nebulae measured via slit spectroscopy \citep[e.g.][]{Lequeux1979, Magrini2009}.

\subsection{Strong Line Metallicity} \label{sec:indirect}
In instances where auroral lines are not detected, empirical calibrations relying on strong line ratios can be used to estimate the gas phase oxygen abundance. Given the wavelength coverage of KCWI at the time of these observations, we do not have access to the \nii6584\ang, \sii6717,6731\ang, and \siii9069,9532\ang \, lines used in many of these calibrations.  
One commonly used calibration --- for which we have all required emission lines --- makes use of the $R_{23}$ emission line ratio proposed by \citet{Pagel1979}:
\begin{equation}
    \small
    \rm R_{23} = \frac{F(\oii3727,3729\AA)+F(\oiii4959\AA)+F(\oiii5007\AA)}{F(H\beta)}
\end{equation}

Unfortunately, this calibration with $R_{23}$ and oxygen abundance is degenerate and therefore requires an additional parameter to break the degeneracy. Various methods of doing this have been proposed with two of the most common being the theoretical calibration from \citet{KK04} (hereafter KK04) and an empirical calibration from \citet{PT05} (PT05). These calibrations rely on an additional line ratio making use of the \oii, and \oiii \, lines in order to provide a tracer of the ionization parameter. Additionally, these methods both have unique calibrations for the ``upper" and ``lower" branches of the $R_{23}$ vs. 12+log(O/H) diagnostic. 
The transition between the two branches lies at $\rm 12+log(O/H)=8.4$ for the \citetalias{KK04} method and there is a ``transition zone" from $\rm 8.0<12+log(O/H)<8.25$ with the \citetalias{PT05} method. The median \hii region metallicity measured for IC\,10 via the direct method is $\rm 12+log(O/H)\sim8.4$, which falls just within the lower branch of the \citetalias{KK04} calibration but the upper branch of the \citetalias{PT05} (with uncertainties overlapping the transition zone). For consistent comparison, the same branch of the two methods should be compared, i.e., lower branch to lower branch.

In the \citetalias{KK04} calibration, the oxygen abundance for each branch is dependent on both $R_{23}$ and the ionization parameter, q. The ionization parameter is also dependent on oxygen abundance and the $O_{32}$ line ratio (Equation \ref{eqn:KK04_O32}), making the solution iterative. Interested readers should refer to Equations 13, 16, and 17 of \citetalias{KK04} for the full polynomial equations of q and oxygen abundance for each branch. In the case of IC\,10, the solution converges on average after 4 iterations for the integrated \hii region spectra and 1-6 iterations for the individual spaxel spectra.
\begin{equation}
    \rm O_{32}=\frac{F(\oiii4959\AA)+F(\oiii5007\AA)}{F(\oii3727,3729\AA)} \\ \label{eqn:KK04_O32}
\end{equation} 

With the \citetalias{PT05} calibration, on the other hand, oxygen abundance is simply a function of the $R_{23}$ and $P$ (Equation \ref{eqn:PT05_p}) emission line ratios. See \citetalias{PT05} Equations 22 \& 24 for the equations for oxygen abundance for each branch.
\begin{equation} \label{eqn:PT05_p}
    \footnotesize
    \rm P= \frac{F(\oiii4959\AA)+F(\oiii5007\AA)}{F(\oiii4959\AA)+F(\oiii5007\AA)+F(\oii3727,3729\AA)} 
\end{equation}

The individual and median oxygen abundance determined by using these strong line calibrations in the integrated \hii region spectra with $\rm SNR_{\oiii4363}>3$ is reported in Table \ref{tab:regions_met} for each branch. The estimates from the \citetalias{PT05} calibration are systematically lower than those determined from the \citetalias{KK04} method on the same branch. This offset between the two methods is well documented with the ``true" value typically assumed to lie between the results of these two calibrations \citep[e.g.,][]{Kewley2008, Moustakas2010}.

\begin{figure}
    \centering
    \includegraphics[width=0.48\textwidth]{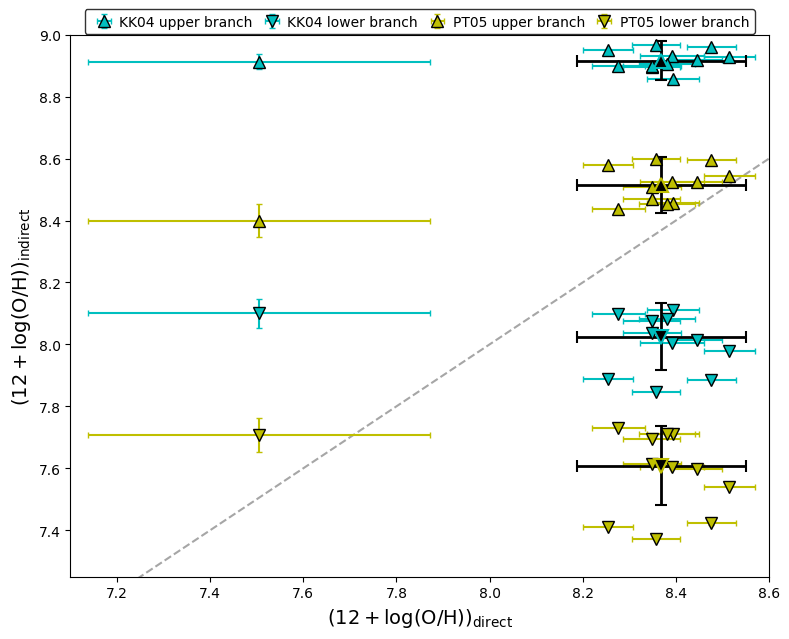}
    \caption{Comparison of the oxygen abundance determined for IC\,10's \hii regions from the direct (horizontal axis) and indirect methods (vertical axis) for regions with $\rm SNR_{\oiii4363}>3$ in the integrated spectrum. The lower branches of the indirect methods are denoted by triangles pointed down and the upper branches by upward pointing triangles. The points corresponding to the \citetalias{KK04} calibration are colored in cyan with \citetalias{PT05} corresponding to yellow points. The filled black points are the medians for each calibration and branch. The dashed grey line indicates the one-to-one line. There is not a clear ``best'' strong line method calibration for which the indirect oxygen abundance matches the value or span of abundances from the direct method. The upper and lower branches bracket the values measured with the direct method regardless of the calibration used, and the span of values probed by each branch/calibration is less than that found with the direct method in all cases.}
    \label{fig:met_comp_integrated}
\end{figure}

As these methods do not require detection of \oiii4363\ang, we can also use them to estimate the oxygen abundance throughout the full field covered in our observations. First, however, we compare the results for the integrated \hii region spectra to those from the direct method. The estimates for each individual \hii region along with the medians are listed in Table \ref{tab:regions_met} and plotted in Figure \ref{fig:met_comp_integrated}. 

As can be seen from the figure and the medians reported in the table, the oxygen abundance determined from the direct method does span a wider range of values than the indirect estimates($\rm 12+log(O/H)\sim 7.5 - 8.5$ versus $\rm0.1-0.4\,dex$), but generally falls in between results of the upper and lower branches. This may be due in part to the proximity of the metallicities being probed to the transition between branches for both calibrations. This makes the usual method of assuming a single branch for estimates of oxygen abundance in a single galaxy introduce a non-negligible systematic uncertainty. Therefore we will not limit our study of the oxygen abundance in individual spaxel spectra to a single branch of the strong line calibrations. The values determined for the oxygen abundance in each \hii region are reported in Table \ref{tab:regions_met}.

\begin{deluxetable*}{lccccc} 
	 \tablecaption{Metallicity Estimates \label{tab:regions_met}} 
\tablehead{\colhead{Region ID} & \colhead{Direct Method} & \colhead{KK04 Upper Branch} & \colhead{KK04 Lower Branch} & \colhead{PT05 Upper Branch} &  \colhead{PT05 Lower Branch} \\ 
	 \colhead{} & \colhead{12+log(O/H)} & \colhead{12+log(O/H)} & \colhead{12+log(O/H)} & \colhead{12+log(O/H)} & \colhead{12+log(O/H)}} 
\startdata 
H16a & 8.36 $\pm$ 0.05 & 8.97 $\pm$ 0.00 & 7.85 $\pm$ 0.00 & 8.60 $\pm$ 0.00 & 7.37 $\pm$ 0.00 \\ 
H16b & 8.35 $\pm$ 0.06 & 8.90 $\pm$ 0.00 & 8.08 $\pm$ 0.00 & 8.47 $\pm$ 0.00 & 7.70 $\pm$ 0.00 \\ 
H17a & 8.28 $\pm$ 0.06 & 8.90 $\pm$ 0.00 & 8.10 $\pm$ 0.00 & 8.44 $\pm$ 0.00 & 7.73 $\pm$ 0.00 \\ 
H18a & 8.39 $\pm$ 0.06 & 8.86 $\pm$ 0.00 & 8.11 $\pm$ 0.00 & 8.46 $\pm$ 0.00 & 7.71 $\pm$ 0.00 \\ 
H18b & 8.45 $\pm$ 0.05 & 8.92 $\pm$ 0.00 & 8.01 $\pm$ 0.00 & 8.52 $\pm$ 0.00 & 7.60 $\pm$ 0.00 \\ 
H18c & 8.39 $\pm$ 0.07 & 8.93 $\pm$ 0.00 & 8.00 $\pm$ 0.00 & 8.53 $\pm$ 0.00 & 7.60 $\pm$ 0.01 \\ 
I18 & 8.38 $\pm$ 0.06 & 8.90 $\pm$ 0.00 & 8.08 $\pm$ 0.00 & 8.45 $\pm$ 0.00 & 7.71 $\pm$ 0.00 \\ 
L11 & 8.48 $\pm$ 0.05 & 8.96 $\pm$ 0.00 & 7.88 $\pm$ 0.00 & 8.60 $\pm$ 0.00 & 7.42 $\pm$ 0.00 \\ 
M11 & 8.51 $\pm$ 0.05 & 8.93 $\pm$ 0.00 & 7.98 $\pm$ 0.00 & 8.54 $\pm$ 0.00 & 7.54 $\pm$ 0.00 \\ 
M12 & 8.25 $\pm$ 0.05 & 8.95 $\pm$ 0.00 & 7.89 $\pm$ 0.00 & 8.58 $\pm$ 0.00 & 7.41 $\pm$ 0.00 \\ 
M16a & 8.35 $\pm$ 0.06 & 8.90 $\pm$ 0.00 & 8.04 $\pm$ 0.00 & 8.51 $\pm$ 0.00 & 7.61 $\pm$ 0.01 \\ 
M16c & 7.51 $\pm$ 0.37 & 8.91 $\pm$ 0.02 & 8.10 $\pm$ 0.05 & 8.40 $\pm$ 0.05 & 7.71 $\pm$ 0.05 \\[0.4em]
Median & 8.37 $\pm$ 0.18 & 8.92 $\pm$ 0.06 & 8.02 $\pm$ 0.11 & 8.52 $\pm$ 0.09 & 7.61 $\pm$ 0.13\\
\enddata 
\tablecomments{Oxygen abundance estimated from the integrated \hii region spectra via the direct (\oiii4363\ang) and strong line ($\rm R_{23}$) methods. Measurements with an uncertainty of 0.00 here denote a measurement uncertainty smaller than two decimal places (this does not include systematic uncertainties discussed in Section \ref{sec:unc}).} 
\end{deluxetable*} 

\subsubsection{Spatially Resolved Abundances}
\begin{figure*}[h]
    \gridline{\fig{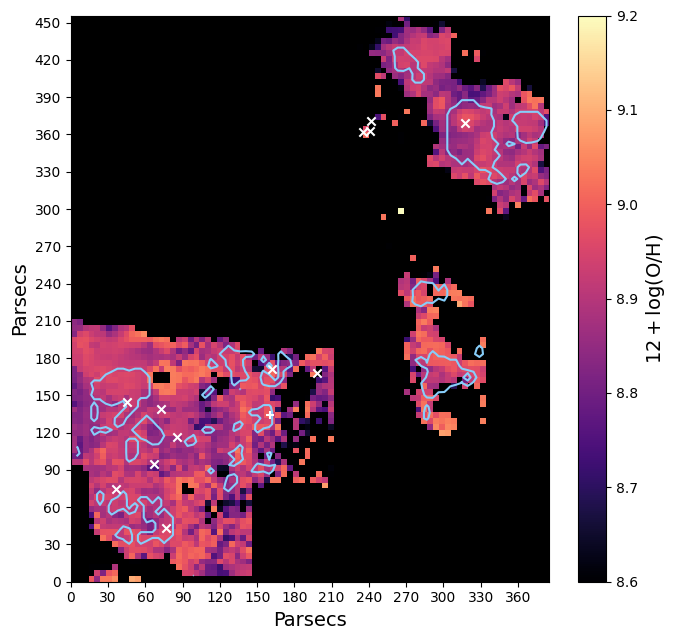}{0.45\textwidth}{(a) KK04, upper branch}
            \fig{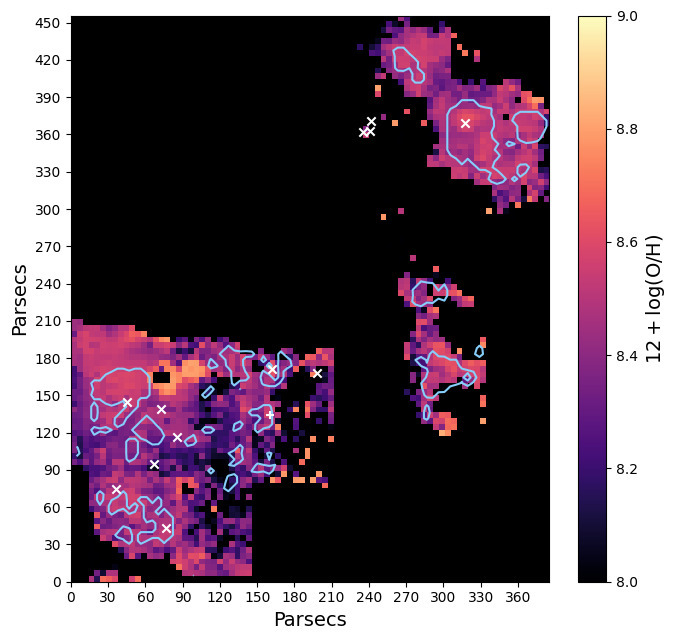}{0.45\textwidth}{(b) PT05, upper branch}}
    \gridline{\fig{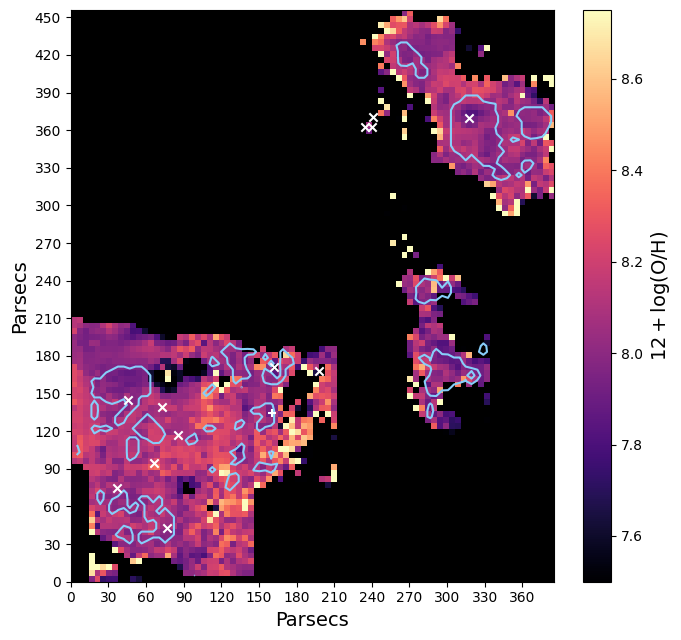}{0.45\textwidth}{(c) KK04, lower branch}
            \fig{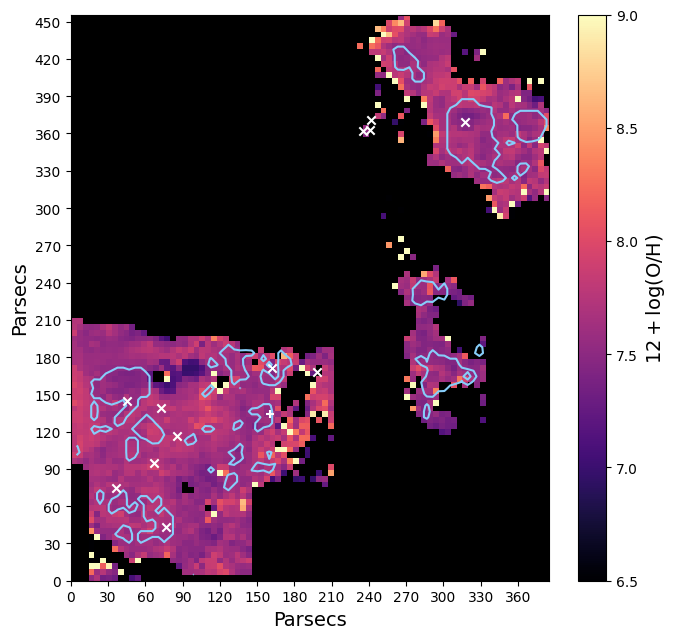}{0.45\textwidth}{(d) PT05, lower branch}}
    \caption{Maps of the estimated oxygen abundance throughout our KCWI observations of IC\,10 using the \citetalias{KK04} (left) and \citetalias{PT05} (right) strong line calibrations. Only spaxels with $\rm SNR_{\oii}>3$ are included. The outlines of the \hii regions identified in \citetalias{Cosens2022} are shown in light blue. The locations of spectroscopically confirmed Wolf-Rayet (WR) stars are shown with white X's and unconfirmed with white $\rm +$'s. Potential chemical enrichment from these WR stars is discussed in Section \ref{sec:WRstars}. These maps show the well-known systematic offset between the \citetalias{KK04} and \citetalias{PT05} calibrations as well as the differing results of the upper and lower branches of the same calibration. The upper branches show higher abundance outside of the \hii regions while the opposite is seen with the lower branches. This illustrates the need for caution in the interpretation of these indirect abundance metrics when applied to the DIG where the physical conditions are likely to be different than in \hii regions for which the calibrations were developed.}
    \label{fig:indirect_metallicity}
\end{figure*}

The analysis of the individual spaxel spectra to produce maps of the estimated oxygen abundance throughout IC\,10 is carried out in the same way as the integrated spectra. Maps of the oxygen abundance produced from both branches of the \citetalias{KK04} and \citetalias{PT05} calibrations are shown in Figure \ref{fig:indirect_metallicity} for spaxels with $\rm SNR_{\oii}>3$. The median oxygen abundance inside and outside of the \hii regions are listed in Table \ref{tab:resolved_metallicity}. The medians for all spaxels inside of \hii regions are well matched to the result of the same calibration applied to the integrated spectra. This is reassuring that the resolved estimates of the oxygen abundance produce physically reasonable results -- at least within an \hii region.

We find small differences between the median oxygen abundance inside and outside the \hii regions, although the direction of the change depends on the calibration branch. Using the lower branch there is a 0.03\,dex (0.02\,dex) higher oxygen abundance outside of the \hii regions for the \citetalias{KK04} (\citetalias{PT05}) method, but the upper branches produce 0.01\,dex (0.06\,dex) lower oxygen abundances. We caution against placing too much emphasis on this and other similar results especially with the sign change depending on the calibration branch. These strong line calibrations are developed based on the physics and observation of star forming regions and may not apply in the DIG where the physical conditions and source of ionization are likely to be different.

In \citetalias{Cosens2022}, we also investigated the resolved abundances via the lower branches of these calibrations for the East portion of the field but with significantly shorter integration times -- only 30s per field compared to 1200s in the current study. With the shallower observations we found an overall higher oxygen abundance than we measure here, with that difference more pronounced outside of the \hii regions (\citetalias{Cosens2022} Figure 22). This is due to lower measured $\rm O_{32}$ ratios resulting primarily from the method of extinction correction. In \citetalias{Cosens2022} there was not sufficient SNR to calculate the extinction on a spaxel-by-spaxel basis as was done here, and instead stacked spectra for all spaxels either inside or outside of an \hii region were used to determine average extinction values in these two regimes. For spaxels outside of \hii regions this resulted in a $\rm 0.17\,dex$ higher extinction value on average over the current study which in turn causes a $\rm \sim1.2\times$ lower $\rm O_{32}$ ratio. This is very close to the measured difference in $\rm O_{32}$ ratios, with the average value outside of the \hii regions being $\rm 1.4\times$ lower in \citetalias{Cosens2022}. The remaining discrepancy is within the measurement uncertainty for the Gaussian fits to the emission lines. The difference in estimated extinction ($\rm 0.03\,dex$) and $\rm O_{32}$ ratio ($\rm 1.1\times$ lower) inside of the \hii regions was much less pronounced, likely due to the higher SNR in those spaxels.

\begin{deluxetable*}{lccc}
     \tablecaption{Strong Line Calibration Results -- Resolved\label{tab:resolved_metallicity}}
     \tablehead{\colhead{Calibration} & \colhead{Median Inside Regions} & \colhead{Median Outside Regions} & \colhead{Total Median} \\
     \colhead{} & \colhead{$\rm 12+log(O/H)$} & \colhead{$\rm 12+log(O/H)$} & \colhead{$\rm 12+log(O/H)$}}
\startdata 
\multicolumn{4}{c}{Lower Branch} \\ 
\tableline 
KK04 & 8.06 $\pm$ 0.01 & 8.09 $\pm$ 0.01 & 8.08 $\pm$ 0.01 \\ 
PT05 & 7.63 $\pm$ 0.01 & 7.65 $\pm$ 0.04 & 7.64 $\pm$ 0.04 \\ 
\tableline 
\multicolumn{4}{c}{Upper Branch} \\ 
\tableline 
KK04 & 8.92 $\pm$ 0.004 & 8.91 $\pm$ 0.01 & 8.91 $\pm$ 0.01 \\ 
PT05 & 8.47 $\pm$ 0.01 & 8.41 $\pm$ 0.01 & 8.43 $\pm$ 0.01 \\
\enddata 
\tablecomments{Median oxygen abundance -- and error on the median -- derived from the \citetalias{KK04} and \citetalias{PT05} strong line calibrations for the upper and lower branches using individual spaxel spectra. The medians are evaluated inside and outside the contours of the identified \hii regions as well as over all spaxels.}
\end{deluxetable*}

\section{Discussion} \label{sec:discussion}

\subsection{ISM Mixing}\label{sec:mixing}
It has been somewhat accepted that the ISM of typical dwarf galaxies is well mixed -- meaning that the oxygen abundances measured across the galaxy are consistent within their uncertainties \citep[e.g.,][]{Berg2012, Haurberg2013}. As large scale IFS observations have only been possible recently, the bulk of the literature in this area makes use of slit spectroscopy in which usually only a few \hii regions are included for each galaxy. Now, with instruments like KCWI and MUSE, this can begin to be evaluated on spatially resolved or semi-resolved scales with many \hii regions probed throughout a single galaxy simultaneously. Even so, most dwarf galaxies appear to be chemically homogeneous \citep[e.g.,][]{Perez-Montero2011, Kumari2019} with only a handful of examples to the contrary \citep[e.g.,][]{James2020}.

Given that spatially resolved measurements of the oxygen abundance have only recently been possible, we will use multiple metrics to determine whether the ISM of IC\,10 is well mixed in order to facilitate comparison with different studies. We will look at (i) the consistency of nominal values within their $\rm 1\sigma$ errors, (ii) the consistency within and size of the $\rm 1\sigma$ distribution, and (iii) the consistency within and size of the interquartile range (IQR). Method (i) has been used in slit spectroscopy studies like \citet{Berg2012} and will facilitate comparison with such multi-galaxy samples. The size and consistency of measurements within the $\rm 1\sigma$ distribution and IQR have been used in resolved studies over individual galaxies \citep[e.g.,][]{James2020, Peng2023} to better capture the deviations within the ISM. Both distributions have been used due to the IQR being less sensitive to outliers, and therefore, potentially better suited to cases where the distribution of oxygen abundance measurements is not Gaussian \citep{James2020}.

\begin{figure*}
    \centering
    \gridline{\fig{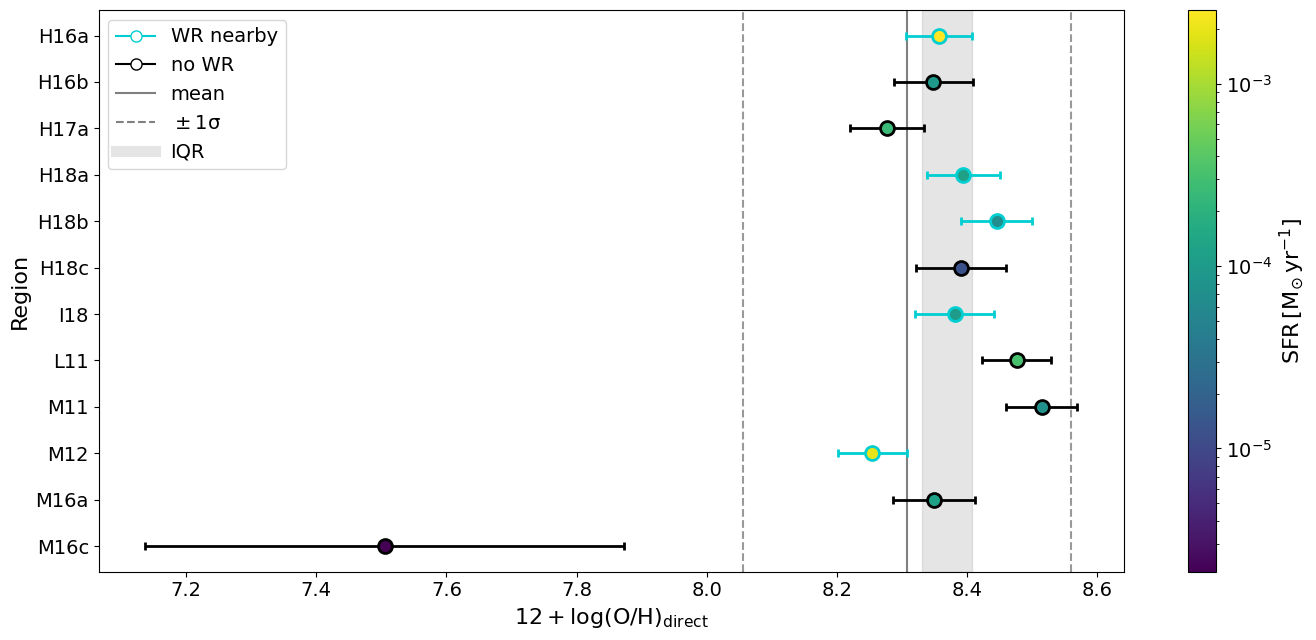}{0.95\textwidth}{(a): All $\rm SNR_{\oiii4363}>3$}}
    \gridline{\fig{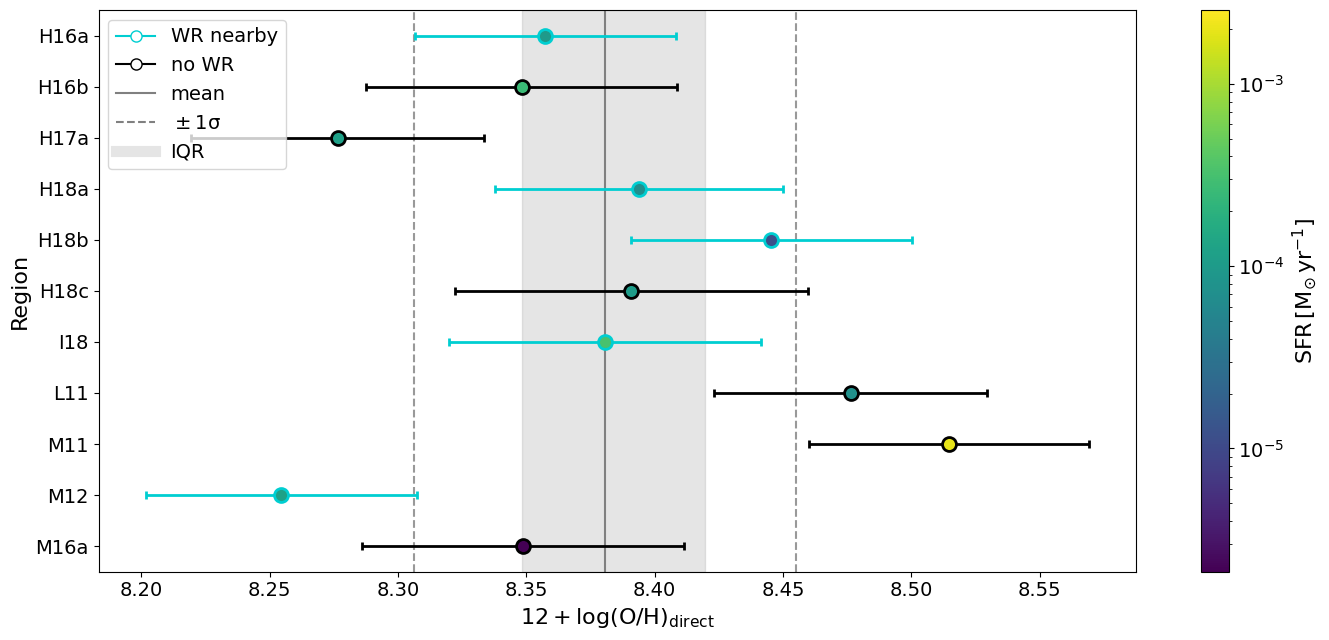}{0.95\textwidth}{(b): Dropped outlier}}
    \caption{Oxygen abundance and uncertainty determined via the direct method for individual \hii regions in IC\,10 where $\rm SNR_{\oiii4363}>3$ (a) and with the outlier region M16c left out of the comparison (b). The data point for each region is filled based on the SFR determined in \citetalias{Cosens2022}, with \hii  regions containing or adjacent to a WR star outlined in cyan. The vertical axis is sorted by region ID. The assignment of these IDs is based on their location, so adjacent regions on this plot are near each other in IC\,10. The mean and $\rm 1\sigma$ deviation of the distribution are shown by the solid and dashed vertical lines. As this distribution is not Gaussian, the IQR is shown by the gray shaded region. As can be seen by the spread of oxygen abundance and how many regions fall outside of the IQR, the ISM throughout IC\,10 may not be well mixed overall, though regions within $\rm\sim100\,pc$ of each other are. This is true even with the outlier M16c left out of the comparison, with 4/11 regions outside of the IQR.}
    \label{fig:ISMmixing}
\end{figure*}

In Figure \ref{fig:ISMmixing}, we plot the measured direct oxygen abundance with uncertainties for each \hii  region for which the $\rm SNR_{\oiii4363}>3$. The data are sorted by \hii region ID which is in turn determined by location in the galaxy (see \citetalias{Cosens2022}) so that adjacent points are in physical proximity.

We find that the measured oxygen abundance within the regions is not consistent within the $\rm 1\sigma$ uncertainties for each region, and that $\rm 1\sigma$ distribution of the sample is rather large, $\pm0.25\,$dex. Only a single region is outside of this range, but the distribution of oxygen abundances is clearly not Gaussian, so we also compare the regions with the smaller IQR of $0.08\,$dex that is less sensitive to outliers like the \hii region M16c. As can be seen by the shaded region in Figure \ref{fig:ISMmixing}, 4/12 of the \hii regions lie fully outside of this range with the $\rm 1\sigma$ error bar of H17a overlapping the IQR by only $0.01\,$dex. This all combines to indicate that the ISM in IC\,10 is, overall, not well mixed. 

M16c is an outlier in not only it's measured abundance, but also in its size and luminosity, being fainter and more compact than the other regions for which we detect \oiii4363\ang. This leads to a lower confidence detection, with M16c as the only region for which $\rm SNR_{\oiii4363}<5$ in the integrated spectrum. Therefore, we repeat this analysis with M16c excluded to further ensure that the results are not biased by this region. These results are still consistent with what we find with all twelve regions included. The $\rm 1\sigma$ distribution shrinks significantly to $\pm0.08\,$dex, but there are still 4/11 regions that fall fully outside of the $0.07\,$dex IQR, which is consistent with the full sample analysis. Regions in close proximity ($\rm \lesssim 100\,pc$) to each other, however, do have measured abundances which are consistent within the uncertainties indicating they are well mixed on local scales, with the exception of the M16 complex and the lower metallicity of M12 relative to L11/M11. The region M12 contains a WNE type WR star and will be discussed further in Section \ref{sec:WRstars}.

The discrepancy between what we find for IC\,10 in this deep, resolved study and what has been seen in the majority of dwarf galaxies could be due to a couple of possible scenarios. First, it could simply be a selection function in previous studies where observations were limited to the brightest regions (slit spectroscopy) or narrower fields (IFS) leading to the appearance of a well mixed ISM. 

Second, we may be observing IC\,10 at a unique point in its evolution. In the handful of dwarf galaxies that have been observed to be inhomogeneous, the explanations for this have varied significantly. This inhomogeneity is sometimes attributed to SNe outflows of enriched gas \citep[blue compact dwarfs Haro\,11 and JKB\,18;][]{James2013a, James2020}, accretion of pristine gas \citep[UM\,448, NGC\,4449;][]{James2013b, Kumari2017}, disruption by a prior or ongoing interaction with a neighbor \citep[UM\,461, J084220+115000;][]{James2010, Fernandez-Arenas2023}, or bursts of star-formation more recent than the mixing timescales \citep[four dwarf spirals;][]{Bresolin2019}.
The conditions in IC\,10 could imply a contribution from multiple of these processes. IC\,10 is in the midst of a starburst which could lead to a temporary metallicity gradients \citep{Bresolin2019}, and it is also posited to be either a late stage merger \citep{Ashley2014} or actively accreting primordial gas \citep{WilcotsMiller1998}. These characteristics could lead to IC\,10 being in a temporary state of having a poorly mixed ISM.

We can also compare these scenarios with simulations of dwarf galaxies. Most of the simulations in the literature have focused on stellar metallicity, but there has been some investigation of the gas phase abundance as well, in particular with the Feedback in Realistic Environments (FIRE) and FIRE-2 cosmological zoom-in simulations. Some studies of dwarf galaxies in these simulation suites have found stellar metallicity gradients, but with a simultaneously well-mixed ISM resulting in a lack of gas-phase abundance gradients \citep[e.g., ][]{Escala2017, Mercado2020}. However, \citet{Porter2022} do find gas-phase abundance gradients in a subset of their sample of five simulated FIRE-2 dwarfs. They find very little gradient in three of the simulated galaxies, but the galaxies with significant metal poor gas accretion and an ongoing major merger display steep gradients in the gas-phase abundances. The authors also note that while one of the galaxies does not exhibit a strong radial abundance gradient, there is a large overall spread in metallicity attributed to widespread starburst episodes.  These scenarios are all consistent with the conditions in IC\,10 discussed above and could therefore contribute to the observed variation in gas-phase oxygen abundance throughout its ISM.

\subsection{Trends with \hii Region Properties}\label{sec:prop_trends}
In \citetalias{Cosens2022}, we measured the kinematic and morphologic properties of these \hii regions at higher spatial and spectral resolution than the observations used here.
To quantitatively test whether these measured \hii region properties (e.g., radius, luminosity, velocity dispersion) in IC\,10 are correlated with the oxygen abundance we evaluate the Spearman rank coefficient, $\rm r_s$. Measurement uncertainties are included by running 10,000 iterations with each data point being perturbed by its associated 1$\sigma$ uncertainty. The median and standard deviation of the resulting distribution of Spearman coefficients and their p-value is reported in Table \ref{tab:spearman_coefficients} for both directly measured and derived properties. 

\begin{figure*}
    \centering
    \includegraphics[width=0.95\textwidth]{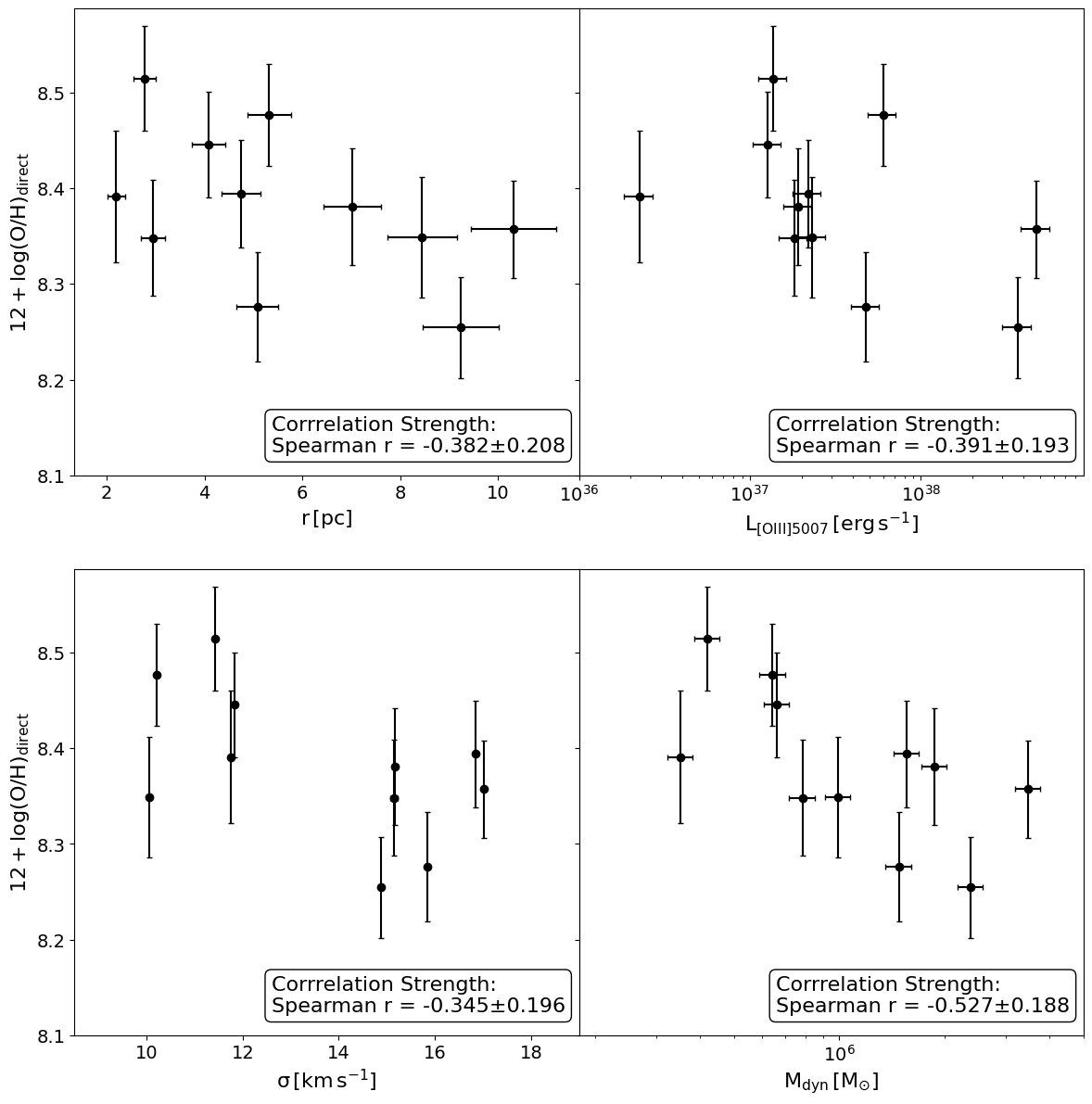}
    \caption{Properties which show evidence of a possible correlation with the gas phase oxygen abundance ($\rm 12+log(O/H)$) when the outlier region M16c is removed. The region radius is shown in the top left, luminosity on the top right, velocity dispersion on the bottom left, and dynamical mass on the bottom right. The Spearman rank coefficients are included on the plots and provide an indication of how likely a correlation between the two parameters is. The strongest (still moderate) correlation is between oxygen abundance and dynamical mass -- though the latter is dependent on $\rm r$ and $\rm \sigma$ as seen in Equation \ref{eqn:Mdyn}.}
    \label{fig:correlated_props}
\end{figure*}

This correlation analysis was performed on the full sample of regions with $\rm SNR_{\oiii4363}>3$, but due to the small sample size, the result is largely influenced by the lower metallicity \hii region, M16c. The analysis is therefore repeated with this outlier region excluded. Both of these iterations are reported in Table \ref{tab:spearman_coefficients}. Those properties which show evidence of a potential correlation ($\rm r_s >0.3$) in the analysis without M16c are highlighted in Figure \ref{fig:correlated_props} with the measured values of the region property reported in Table \ref{tab:reg_props}. For full \hii region properties we refer the interested reader to \citetalias{Cosens2022}.

There are weak correlations between the oxygen abundance and the \hii region radius ($\rm r_s=-0.38\pm 0.21$), \oiii5007\ang \, luminosity ($\rm r_s=-0.39\pm 0.20$), and velocity dispersion ($\rm r_s=-0.35\pm 0.20$). Given the p-values for these comparisons are around $\rm0.25\pm0.25$ (and the large uncertainties on $\rm r_s$), there is a possibility that these same Spearman rank coefficients could be achieved with data from independent distributions. There are also weak correlations with the ionized gas mass, $\rm M_{\hii}$, and outward pressure from turbulence, $\rm P_{turb}$, but these are derived properties and are therefore excluded in Figure \ref{fig:correlated_props}. The strongest trend observed ($\rm r_s=-0.53\pm 0.19$) is between oxygen abundance and the dynamical mass estimate, $\rm M_{dyn}$, however it should be noted that this is dependent on both the region size and velocity dispersion:
\begin{equation}
    \rm M_{dyn} = \frac{5\sigma^2r}{G} \label{eqn:Mdyn}
\end{equation}

There are some key differences between what we find here and what has been found in previous studies. A notable example is the PHANGS collaborations study of eight nearby spiral galaxies observed with MUSE \citep{Kreckel2019}. They find a negative correlation between (indirect measurements of) oxygen abundance and velocity dispersion, but positive correlations with size and luminosity. This is in contrast to the negative correlations we see between oxygen abundance and all three of these parameters in the \hii regions of IC\,10, though these are weak correlations and therefore cannot rule out consistency with the \citet{Kreckel2019} sample. Still, one might expect to see different trends due to fundamental differences in the characteristics of the galaxies measured  --- spiral galaxies with clear radial abundance gradients versus a poorly mixed dwarf --- and the spatial resolution and extent of the observations --- $\rm\sim50\,pc$ over the full disk versus $\rm\sim5\,pc$ over the central region. As the sample of galaxies (dwarf and spiral) studied with IFS increases these correlations can be studied with more certainty than in IC\,10 to answer whether there are environmental dependencies and learn more about the mechanisms responsible for driving gas mixing in the ISM.

\begin{deluxetable*}{lcccc}
     \tablecaption{Spearman Coefficients -- Relative to 12+log(O/H)\label{tab:spearman_coefficients}}
     \tablehead{\colhead{Property} & \colhead{$\rm r_s$} & \colhead{p} & \colhead{$\rm r_s$} & \colhead{p} \\
     \colhead{} & \multicolumn{2}{c}{All regions} & \multicolumn{2}{c}{Dropped outlier}}
\startdata 
Radius, $\rm r$ & $-$0.14 $\pm$ 0.18 & 0.62 $\pm$ 0.25 & $-$0.38 $\pm$ 0.21 & 0.25 $\pm$ 0.26 \\
Luminosity, $\rm L_{\oiii5007}$ & $-$0.07 $\pm$ 0.15 & 0.71 $\pm$ 0.20 & $-$0.39 $\pm$ 0.19 & 0.23 $\pm$ 0.25 \\
Velocity Dispersion, $\rm \sigma_{\oiii5007}$ & $-$0.04 $\pm$ 0.15 & 0.73 $\pm$ 0.20 & $-$0.35 $\pm$ 0.20 & 0.30 $\pm$ 0.26 \\
Dynamical Mass, $\rm M_{dyn}$ & $-$0.18 $\pm$ 0.15 & 0.57 $\pm$ 0.24 & $-$0.53 $\pm$ 0.19 & 0.10 $\pm$ 0.19 \\
Ionized Gas Mass, $\rm M_{\hii}$ & $-$0.07 $\pm$ 0.16 & 0.70 $\pm$ 0.21 & $-$0.39 $\pm$ 0.20 & 0.23 $\pm$ 0.26 \\
SFR Surface Density, $\rm \Sigma_{SFR}$ & 0.09 $\pm$ 0.17 & 0.68 $\pm$ 0.23 & $-$0.18 $\pm$ 0.21 & 0.56 $\pm$ 0.27 \\
Extinction, $\rm E(B-V)$ & $-$0.19 $\pm$ 0.16 & 0.54 $\pm$ 0.25 & 0.05 $\pm$ 0.19 & 0.69 $\pm$ 0.23 \\
Radiation Pressure, $\rm P_{dir}$ & 0.09 $\pm$ 0.17 & 0.70 $\pm$ 0.23 & $-$0.17 $\pm$ 0.21 & 0.56 $\pm$ 0.27 \\
Warm Gas Pressure, $\rm P_{gas}$ & 0.23 $\pm$ 0.19 & 0.46 $\pm$ 0.28 & $-$0.01 $\pm$ 0.25 & 0.61 $\pm$ 0.26 \\
Turbulent Pressure, $\rm P_{turb}$ & \nodata & \nodata & $-$0.33 $\pm$ 0.28 & 0.39 $\pm$ 0.29 \\
\oii3727\ang/\hb & $-$0.33 $\pm$ 0.15 & 0.30 $\pm$ 0.22 & $-$0.13 $\pm$ 0.18 & 0.65 $\pm$ 0.25 \\
\enddata 
\tablecomments{Spearman rank coefficients and associated p-values determined between oxygen abundance and \hii region properties. It should be noted that the p-values cannot be negative even if the uncertainties would imply a negative value. These coefficients are reported for all regions in which a direct metallicity could be estimated and for the case in which the outlier region M16c is excluded. There is a significant difference in the results for these two cases except for the case of $\rm P_{turb}$ since the value could not be computed for M16c due to the measured velocity dispersion being less than the estimated sound speed for the region.}
\end{deluxetable*}

\subsubsection{Ionizing Photon Escape Fraction}
Previous studies of the ionizing photon escape fraction, $\rm f_{esc}$, have found trends between this parameter and the luminosity of the \hii region \citep[e.g.,][]{Pellegrini2012, Teh2023}. The direction of this trend though has differed, with \citet{Teh2023} finding a negative correlation between $\rm f_{esc}$ and \lha \, in NGC\,628 in contrast to \citet{Pellegrini2012} observing more luminous regions in the SMC/LMC to have a larger $\rm f_{esc}$. This is likely due to the dependence of $\rm f_{esc}$ on the environmental conditions in the galaxy such as metallicity \citep{Rahner2017}.

\begin{figure*}
    \centering
    \includegraphics[width=0.99\textwidth]{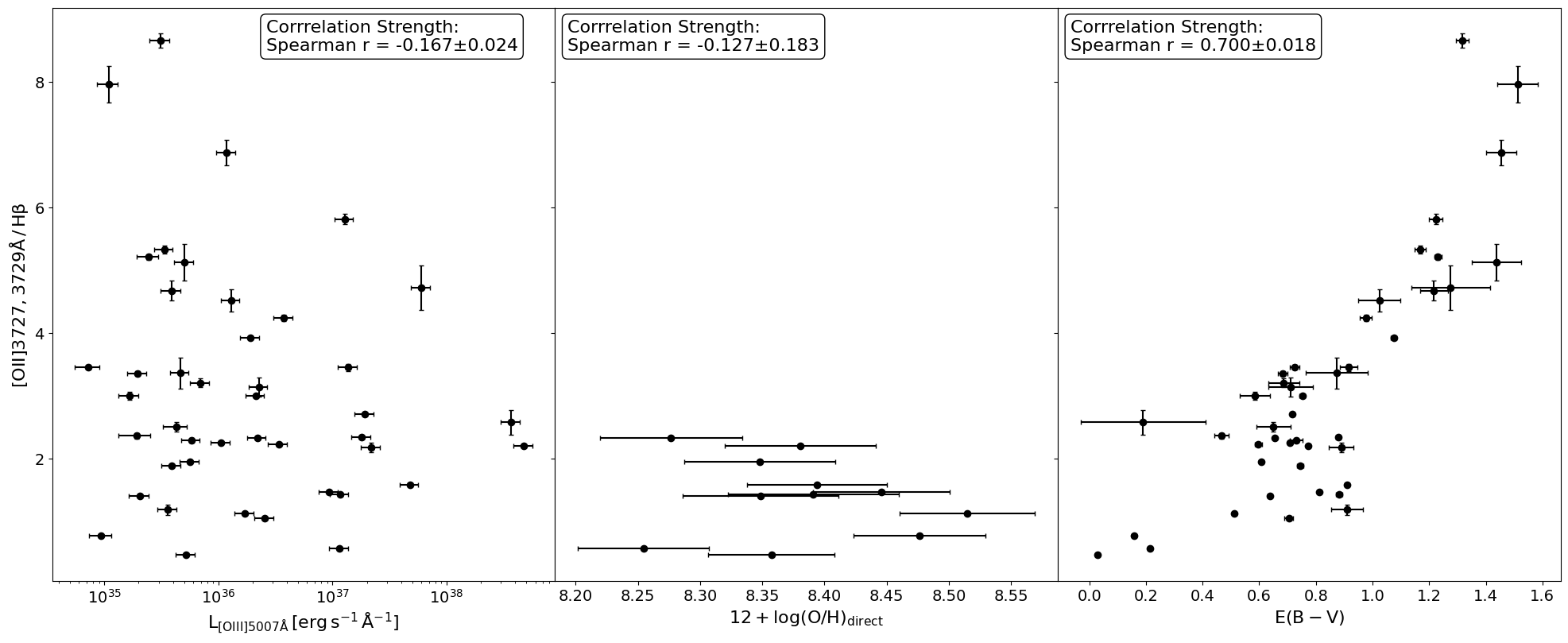}
    \caption{\oii /\hb \, ratio as a function of \oiii5007\ang \, luminosity (left), 12+log(O/H) determined via the direct method (center), and extinction measured from the H$\gamma$/\hb \, line ratio (right). Lower $\rm L_{[OIII]}$ and $\rm 12+log(O/H)$ are weakly correlated with increasing \oii /\hb, but higher extinction is quite strongly correlated with increasing \oii /\hb. Fewer \hii regions are included in the center plot due to requirement that $\rm SNR_{\oiii4363}>3$.}
    \label{fig:O2_Hb}
\end{figure*}

We cannot measure $\rm f_{esc}$ directly in our KCWI data, but \citet{Teh2023} investigated possible correlations between $\rm f_{esc}$ and other properties of the \hii regions in the spectra of NGC\,628. They find a potential correlation between increasing $\rm f_{esc}$ with increasing \oii /\hb \, ratio ($\rm r_s = 0.48$), which we can use as a proxy for $\rm f_{esc}$ in order to check for a trend with \hii region luminosity in IC\,10. While \citet{Teh2023} note that the physical origin of this correlation is unclear, they speculate that it may be due to the ionization parameter or metallicity. Since the \oii /\hb \, ratios in IC\,10 are in the same range as NGC\,628 and the gas phase oxygen abundance is also similar to previous measurements by \citet{Berg2013}, we would expect to be probing a similar regime. The measured value of the \oii /\hb \, line ratio in IC\,10's \hii regions is included in Table \ref{tab:reg_props}. As shown in Figure \ref{fig:O2_Hb}, we find a weak correlation between decreasing $\rm L_{[OIII]5007}$ and increasing \oii /\hb. This indicates that there may be a negative correlation (although a weak one; $\rm r_s = -0.17 \pm 0.02$) in IC\,10 between the region luminosity and $\rm f_{esc}$ similar to NGC\,628.

Further, we find another weak negative correlation ($\rm r_s = -0.13 \pm 0.18$) between the \oii /\hb \, ratio and the oxygen abundance of the region, though we caution that this is very tentative as the p-value is large. The negative sign of this correlation indicates that the \oii /\hb \, ratio decreases with increasing $\rm 12+log(O/H)$. If the trend found by \citet{Teh2023} holds, and the \oii /\hb \, ratio is indeed positively correlated with $\rm f_{esc}$, then this indicates that we are seeing higher $\rm f_{esc}$ in \hii regions with lower oxygen abundance in IC\,10. This would not be an unexpected result, as simulations of star-forming clouds like those in \citet{Kimm2019} find the same result.

However, we also find an increased \oii /\hb \, ratio with measured higher extinction ($\rm r_s = 0.70 \pm 0.02$). This is counter-intuitive as we would expect a more heavily extincted region to exhibit lower $\rm f_{esc}$. As this correlation does not depend on the fainter \oiii4363\ang \, line, it does not suffer from the same effects of small sample size and sensitivity to outliers. Combined with the weak anti-correlation with $\rm 12+log(O/H)$, it would imply an increase in extinction with \textit{lower} oxygen abundance; counter to the trend observed in local galaxy samples \citep[e.g.,][]{Xiao2012}. Further, we do not find any evidence of this trend when directly comparing the two parameters in regions where the auroral \oiii4363\ang \, line is detected (Table \ref{tab:spearman_coefficients}). 
\citet{Teh2023} note that, due to the lack of infrared observations available, their definition of $\rm f_{esc}$ also could include photons absorbed by dust grains. This could explain the positive correlation seen in IC\,10 between the \oii / \hb \, ratio as an $\rm f_{esc}$ proxy and measured extinction. As \citet{Teh2023} is a pilot study of a single galaxy, it will be interesting to see if this correlation is found in a larger sample probing a larger parameter space and if the \oii/\hb \, ratio can be reliably used as an $\rm f_{esc}$ proxy.

\begin{deluxetable*}{ccccccc} 
	 \tablecaption{\hii Region Properties \label{tab:reg_props}} 
\tablehead{\colhead{Region} & \colhead{$\rm r$} & \colhead{$\rm \sigma_{\oiii5007}$} & \colhead{$\rm L_{[OIII]5007}$} & \colhead{$\rm M_{dyn}$} & \colhead{\oii/\hb} & \colhead{$
\rm E(B-V)$}\\ 
	 \colhead{} & \colhead{(pc)} & \colhead{$\rm km \, s^{-1}$} & \colhead{($\rm 10^{35} \,erg \, s^{-1}$)} & \colhead{$\rm 10^5 \, M_{\odot}$} & \colhead{} & \colhead{(mag)}}
\startdata 
H16a & 10.32 $\pm$ 0.87 & 17.02 $\pm$ 0.00 & 4741.00 $\pm$ 889.10  & 34.72 $\pm$ 2.93 & 0.46 $\pm$ 0.00 & 0.03 $\pm$ 0.00 \\ 
H16b & 2.95 $\pm$ 0.25 & 15.15 $\pm$ 0.08 & 181.00 $\pm$ 33.95  & 7.87 $\pm$ 0.67 & 1.95 $\pm$ 0.01 & 0.61 $\pm$ 0.01 \\ 
H17a & 5.08 $\pm$ 0.43 & 15.85 $\pm$ 0.01 & 476.90 $\pm$ 89.41  & 14.82 $\pm$ 1.25 & 2.32 $\pm$ 0.00 & 0.65 $\pm$ 0.00 \\ 
H18a & 4.74 $\pm$ 0.40 & 16.84 $\pm$ 0.03 & 218.70 $\pm$ 41.01  & 15.61 $\pm$ 1.32 & 1.57 $\pm$ 0.01 & 0.91 $\pm$ 0.01 \\ 
H18b & 4.08 $\pm$ 0.34 & 11.83 $\pm$ 0.01 & 127.50 $\pm$ 23.90  & 6.63 $\pm$ 0.55 & 1.46 $\pm$ 0.01 & 0.81 $\pm$ 0.00 \\ 
H18c & 2.19 $\pm$ 0.18 & 11.75 $\pm$ 0.03 & 22.63 $\pm$ 4.24  & 3.51 $\pm$ 0.29 & 1.43 $\pm$ 0.01 & 0.88 $\pm$ 0.01 \\ 
I18 & 7.02 $\pm$ 0.59 & 15.16 $\pm$ 0.06 & 192.30 $\pm$ 36.07  & 18.75 $\pm$ 1.58 & 2.20 $\pm$ 0.01 & 0.77 $\pm$ 0.00 \\ 
L11 & 5.32 $\pm$ 0.45 & 10.21 $\pm$ 0.01 & 601.40 $\pm$ 112.80  & 6.45 $\pm$ 0.55 & 0.76 $\pm$ 0.00 & 0.16 $\pm$ 0.00 \\ 
M11 & 2.77 $\pm$ 0.23 & 11.42 $\pm$ 0.02 & 137.30 $\pm$ 25.75  & 4.20 $\pm$ 0.35 & 1.12 $\pm$ 0.01 & 0.51 $\pm$ 0.00 \\ 
M12 & 9.24 $\pm$ 0.78 & 14.88 $\pm$ 0.00 & 3693.00 $\pm$ 692.40  & 23.78 $\pm$ 2.01 & 0.57 $\pm$ 0.00 & 0.21 $\pm$ 0.00 \\ 
M16a & 8.45 $\pm$ 0.71 & 10.06 $\pm$ 0.02 & 230.30 $\pm$ 43.19  & 9.93 $\pm$ 0.83 & 1.40 $\pm$ 0.01 & 0.64 $\pm$ 0.01 \\ 
M16c & 2.61 $\pm$ 0.22 & 8.91 $\pm$ 0.21 & 3.92 $\pm$ 0.74  & 2.41 $\pm$ 0.22 & 4.71 $\pm$ 0.35 & 1.28 $\pm$ 0.14 \\ 
\enddata 
\tablecomments{Properties of \hii regions in the central region of IC\,10 with $\rm SNR_{\oiii4363}>3$. The values for region $\rm r$, $\rm \sigma_{\oiii5007}$, $\rm L_{\oiii5007}$, and $\rm M_{dyn}$ were determined from the high resolution data in \citetalias{Cosens2022}.} 
\end{deluxetable*} 

\subsection{Impact of Wolf-Rayet Stars} \label{sec:WRstars}
IC\,10 is home to an unusually high number of Wolf-Rayet (WR) stars for its mass and size. These WRs are evolved, massive stars which have transitioned from the hydrogen to primarily helium burning phase with strong stellar winds, stripping them of their hydrogen envelopes. These winds carry material from the star and lead to observed changes in the chemical content of the surrounding ionized gas, with studies finding enhanced nitrogen content and decreased O/H ratios \citep[e.g.,][]{Esteban1992, LopezSanchez2007, LopezSanchez2011}.

Looking at variation in the oxygen abundance of \hii regions that are in close proximity, we see that the $\rm 12+log(O/H)$ is significantly lower in the \hii region M12 compared to its neighbors L11 and M11. Figure \ref{fig:ISMmixing} shows the $\rm >0.2\,dex$ lower abundance in region M12 which hosts a spectroscopically confirmed WNE (an early-type nitrogen sequence WR) star. These regions are distinct, but are in close proximity with $\rm <50pc$ separating them. This indicates that it is possible the WNEs hydrogen envelope is leading to a lower measured O/H ratio in the integrated \hii region spectrum. Deeper observations in which the direct oxygen abundance could be measured on a resolved basis would be needed to confirm this.

\begin{figure}
    \centering
    \includegraphics[width=.49\textwidth]{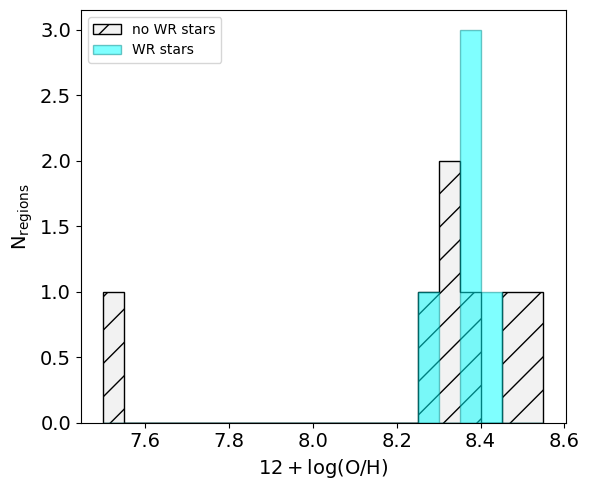}
    \caption{Histogram of direct metallicity for each \hii region in IC\,10 with $\rm SNR_{\oiii4363}>3$. The \hii regions associated with a known or newly detected candidate WR star are shown in cyan while those without a nearby WR star are shown by the black hatched region. There does not appear to be any clear correlation between \hii regions associated with a WR star and gas-phase oxygen abundance.}
    \label{fig:metallicity_WRnearby}
\end{figure}

To test the global impact of the WR winds on the surrounding \hii regions in IC\,10, we compare the oxygen abundance measured in Section \ref{sec:direct} from the direct method for all \hii regions with and without an associated WR star in Figure \ref{fig:metallicity_WRnearby}. In order for a WR star to be considered associated with an \hii region it must be $<$2 pixels ($\rm\sim9\,pc$) away --- though all are contained within the region aside from the WR star R2 associated with region H18b. For this analysis we include known WR stars as well as a candidate WN star detected in our KCWI spectra associated with \hii region H18a (to be discussed further in \citet{Cosens2024b}). We do not find evidence of a trend with metallicity and WR star association though we caution that this is too small of a sample from which to make large scale conclusions. Deeper, and higher spatial resolution observations in the vicinity of the WR stars could probe the extent of their impact in more detail, but this is beyond the scope of the current study.

\subsection{Sources of Uncertainty}\label{sec:unc}
There is one outlier in our sample, M16c, which has an oxygen abundance $\rm 0.77\,dex$ lower than the next closest region in the sample. As noted in Section \ref{sec:mixing}, M16c is fainter and more compact the the other \hii regions with a significantly noisier spectrum. This increase in noise results in a $\rm1\sigma$ uncertainty on the oxygen abundance that is $\rm \sim7\times$ greater than for the other regions. To avoid conclusions which are heavily influenced by one region the analysis of the preceding sections was performed both with and without this region included. The conclusions about ISM mixing in Section \ref{sec:mixing} are the same in each case as the metrics used are flexible in the treatment of outliers. In the investigation of trends between oxygen abundance and \hii region properties of Section \ref{sec:prop_trends}, however, this outlier has a significant impact on the results even with the larger uncertainty due to the high degree of separation from the other values. Therefore we consider the correlation coefficients more reliable with M16c excluded as they do not depend strongly on any other single region -- though the results of both analyses are reported in Table \ref{tab:spearman_coefficients}.

There are also sources of systematic uncertainty which affect the entire sample considered here. Even though we are using the ``direct method" of calculating oxygen abundance from collisionally excited auroral lines, there are underlying assumptions being made. First, due to the wavelength coverage and spectral resolution of our KCWI spectra we are only able to calculate the abundance of the $\rm O^{2+}$ ion and are forced to rely on an empirical relation with the \oiii \, electron temperature for the total abundance. This introduces a systematic uncertainty due to the 0.15\,dex of scatter in this relationship \citep{Amorin2015}.

Additionally, there may be temperature inhomogeneity even within a single ionization zone of an \hii region. This could cause the value of $\rm T_e$ derived in Equation \ref{eqn:Te} to be biased towards higher temperatures than the true average \citep[e.g.,][]{Kewley2019}. \citet{Mendez-Delgado2023} find that this effect is more pronounced in the high-ionization zones probed by \oiii4363\ang \, and can lead to an underestimation of the oxygen abundance, particularly in regions of low metallicity ($\rm 12+log(O/H)\le8$). As this is near the average value found for IC\,10's \hii regions in this sample, we may be slightly underestimating the true value of the oxygen abundance. This could also add to the observation that the gas in IC\,10 is not as well mixed as typical dwarf galaxies due to the possible underestimation being more pronounced for the lower metallicity \hii regions. Unfortunately, we cannot calculate the degree of temperature variation in our current KCWI spectra since the \oiii4363\ang \, line is not sufficiently detected in individual spaxel spectra. 

For our estimation of the oxygen abundance using the strong line method, there are yet additional uncertainties to keep in mind. For empirical calibrations, like the \citetalias{PT05} calibration used here, they are determined based on relationships between the strong line ratios and the auroral line oxygen abundances, so all the uncertainties in the direct method will be carried into the strong line estimates. There can also be very large differences between different calibrations \citep{Kewley2019}, indicating that there is a dependence on the underlying samples used to generate these calibrations that should not be discounted.
Theoretical strong line calibrations, like the \citetalias{KK04} method, can span a wider range of parameter space than empirical calibrations, but assumptions in the modeling can also introduce systematic uncertainties.

For both of these methods, there is an additional uncertainty that is especially important in the case of IC\,10. Both calibrations rely on the $\rm R_{23}$ parameter which is double valued with oxygen abundance and requires the selection of either the ``upper" or ``lower" branch to estimate the oxygen abundance. For IC\,10, however, the average oxygen abundance measured throughout the galaxy in previous studies, as well as with the direct method here, is very near the transition between these branches, meaning there are likely some regions for which the lower branch should be used and others which should use the upper branch. This may be the reason for the lower degree of variation in oxygen abundance seen with the strong line calibrations in comparison to the direct estimates for IC\,10's \hii regions in Figure \ref{fig:met_comp_integrated}.

\section{Summary} \label{sec:conclusion}
We used the Keck/KCWI optical IFS to study the ionized gas conditions in and around the \hii regions of the nearby starburst galaxy IC\,10. In a previous study we used the high-resolution ($\rm 0.35''$ spatial sampling and $\rm R\sim18,000$) mode to study the morphology and kinematics of these regions, while here we use the coarser sampling (both spatial and spectral) and wider wavelength coverage mode of KCWI in order to study the gas phase oxygen abundance throughout these regions and the DIG. We cover $\rm\sim1.7~sq.~arcmin$ with $\rm 1.35''$ spatial sampling, $\rm R\sim900$, and wavelength coverage $\rm3500-5500$\ang. Our main findings with these observations are as follows.
\begin{enumerate}
    \item We capture 43/46 \hii regions that were identified in the high resolution observations, with the auroral \oiii4363\ang \, line detected at $\rm SNR>3$ in 12/43. Using this auroral line, we find that the \hii regions in the central portion of IC\,10 have a median gas phase oxygen abundance of $\rm12+log(O/H)=8.37\pm 0.18$.

    \item Even covering just the central starburst of IC\,10, we find a wide range of oxygen abundance throughout the \hii regions from the direct method ($\rm 7.5 < 12+log(O/H) < 8.5$). By multiple metrics used throughout the literature --- size of/consistency with the $\rm1\sigma$ distribution and IQR, overlap within $\rm1\sigma$ uncertainties --- we find that the ISM in IC\,10 is not well mixed. This is unusual for observations of dwarf galaxies and may be due to either the starburst nature of IC\,10 or possibly to selection biases of previous observations.

    \item We investigate potential trends between the oxygen abundance and other observed properties of the \hii regions. We find that increasing oxygen abundance has weak negative correlations with region $\rm r$, $\rm L_{\oiii5007}$, $\rm \sigma$, and the \oii/\hb \, ratio --- a possible proxy for $\rm f_{esc}$. There is also a moderate negative correlation with the region $\rm M_{dyn}$, itself a product of $\rm r$ and $\rm \sigma$.
    
    \item We compare the oxygen abundance determined via the direct method and the \citetalias{PT05} and \citetalias{KK04} strong line calibrations in the integrated \hii region spectra. We find that these calibrations span a narrower range of oxygen abundances than using the auroral \oiii4363\ang \, line. This may be due to IC\,10 falling near the transition between branches for each of these calibrations, making it difficult to determine which branch gives the ``correct" result for regions which do not have the auroral line detected. 

    \item We use the strong line calibrations to estimate the oxygen abundance throughout the remainder of IC\,10 on a per spaxel basis. We find good agreement between these resolved abundances and the integrated region estimates for each calibration and branch, but caution that outside of the \hii regions there is a high degree of uncertainty with the trend observed being dependent on the branch used. This may be partly due to these spaxels probing a larger degree of the DIG which is not the regime in which the strong line calibrations were derived.
\end{enumerate}

This detailed look at the ionized gas conditions throughout the central starburst of the dwarf galaxy IC\,10 has revealed more variation than is typically seen in dwarf galaxies. This may be partly due to the starburst nature of IC\,10, cold gas accretion, and/or a late stage merger, but it may also be an effect of observing a large portion of the galaxy with the resolution and sensitivity afforded with KCWI. More comprehensive IFS studies of the ionized gas conditions of nearby dwarf galaxies are needed to determine if IC\,10 and the few other dwarfs with poorly mixed ISM are outliers or if we need to revisit the conventional wisdom that dwarf galaxies in general are chemically well-mixed.

\begin{acknowledgments}
M.C.~is supported by a Brinson Prize Fellowship at Carnegie Observatories. The data presented herein were obtained at the W.~M.~Keck Observatory, which is operated as a scientific partnership among the California Institute of Technology, the University of California and the National Aeronautics and Space Administration. The Observatory was made possible by the generous financial support of the W. M. Keck Foundation. The authors wish to recognize and acknowledge the very significant cultural role and reverence that the summit of Maunakea has always had within the indigenous Hawaiian community.  We are most fortunate to have the opportunity to conduct observations from this mountain. We also thank the anonymous referee for their time reviewing this paper and providing helpful feedback.
\end{acknowledgments}

\facilities{Keck:II(KCWI)}

\software{astrodendro \citep{astrodendro},
          Astropy \citep{astropy:2013, astropy:2018}, 
          IPython \citep{ipython}
          Matplotlib \citep{Hunter2007},
          NumPy \citep{numpy},
          pandas \citep{pandas, pandas2},
          photutils \citep{photutils},
          reproject \citep{reproject}
          }

All the {\it HST} data used in this paper  were obtained from the Mikulski Archive for Space Telescopes (MAST) at the Space Telescope Science Institute. The specific observations analyzed can be accessed via \dataset[10.17909/7shz-n457]{http://dx.doi.org/10.17909/7shz-n457}

\bibliography{IC10_BL}{}
\bibliographystyle{aasjournal}

\end{document}